\documentclass[prd,aps,nofootinbib,superscriptaddress,floatfix,preprintnumbers,twocolumn]{revtex4}
\usepackage{hyperref,amssymb,amsmath,graphicx,xcolor,slashed}

\DeclareMathOperator\Tr{Tr}

\begin{document}
\preprint{MSUHEP-20-008}
%%%%%%%%%%%%%%%%%%%%%%%%%%%%%%%%%%%%%%%%%%%%%%%%%%%%%%%%%%%%%%%%%%%%%%%%%%%%%%%%
\title{Probing nucleon strange and charm distributions with lattice QCD}

\author{Rui Zhang}
\affiliation{Department of Physics and Astronomy, Michigan State University, East Lansing, MI 48824}
\affiliation{Department of Computational Mathematics,
  Science and Engineering, Michigan State University, East Lansing, MI 48824}

\author{Huey-Wen Lin}
\email{hwlin@pa.msu.edu}
\affiliation{Department of Physics and Astronomy, Michigan State University, East Lansing, MI 48824}
\affiliation{Department of Computational Mathematics,
  Science and Engineering, Michigan State University, East Lansing, MI 48824}

\author{Boram Yoon}
\affiliation{Computer, Computational, and Statistical Sciences CCS-7, Los Alamos National Laboratory, Los Alamos, NM 87545, USA}

%%%%%%%%%%%%%%%%%%%%%%%%%%%%%%%%%%%%%%%%%%%%%%%%%%%%%%%%%%%%%%%%%%%%%%%%%%%%%%%%
\begin{abstract}
We present the first lattice-QCD calculation of the unpolarized strange and charm parton distribution functions using large-momentum effective theory (LaMET).
We use a lattice ensemble with 2+1+1 flavors of highly improved staggered quarks (HISQ) generated by MILC collaboration, with lattice spacing $a\approx 0.12$~fm and $M_\pi \approx 310$~MeV, and clover valence fermions with two valence pion masses: 310 and 690~MeV. 
We use momentum-smeared sources to improve the signal up to  nucleon boost momentum $P_z=2.18$~GeV, 
and determine nonperturbative renormalization factors in RI/MOM scheme.  
We compare our lattice results with the matrix elements obtained from matching the PDFs from CT18NNLO and NNPDF3.1NNLO global fits. Our data support the assumptions of strange-antistrange and charm-anticharm symmetry that are commonly used in global PDF fits, and we find smaller than expected parton distribution at mid to small $x$. 
\end{abstract}

\maketitle

%%%%%%%%%%%%%%%%%%%%%%%%%%%%%%%%%%%%%%%%%%%%%%%%%%%%%%%%%%%%%%%%%%%%%%%%%%%%%%%%
Parton distribution functions (PDFs) provide a universal description of hadronic constituents as well as critical inputs for the discovery of the Higgs boson found at the Large Hadron Collider (LHC)
through proton-proton collisions~\cite{CMS:2012nga,ATLAS:2012oga}.
While the world waits for the next phase of LHC discovery focused on searching for new-physics signatures, improvements in the precision with which we know Standard-Model backgrounds will be crucial to discern these signals. 
For example, our knowledge of many Higgs-production cross sections remains dominated by PDF uncertainties. 
Among the known PDFs, the strange and charm PDFs have particularly large uncertainty despite decades of experimental effort.
In addition to their applications to the energy frontier, PDFs also reveal a nontrivial structure inside the nucleon, such as its momentum and spin distributions.
Many ongoing and planned experiments at facilities around the world, such as Brookhaven and Jefferson Laboratory in the United States, GSI in Germany, J-PARC in Japan, or a future electric-ion collider (EIC), are set to explore the less-known kinematics of nucleon structure and more. 

In order to distinguish the flavor content (strange or charm) of the PDFs, experiments use nuclear data, such as neutrino scattering off heavy nuclei, and the current understanding of medium corrections in these cases is limited.
Thus, the uncertainty in the strange PDFs remains large.
In many cases, the assumptions $\overline{s}(x)=s(x)$ and $\overline{c}(x)=c(x)$ that are often made in global analyses can agree with the data merely due to the large uncertainty. 
At the LHC, strangeness can be extracted through the $W+c$ associated-production channel, but their results are rather puzzling. 
or example, ATLAS got the ratios of averaged strange and antistrange to the twice antidown distribution, $(s+\overline{s})/(2\overline{d})$, to be $0.96^{+0.26}_{-0.30}$ at $Q^2=1.9\text{ GeV}^2$ and $x=0.023$~\cite{Aad:2014xca}. 
CMS performed a global analysis with deep-inelastic scattering (DIS) data and the muon-charge asymmetry in $W$ production at the LHC to extract the ratios of the total integral of strange and antistrange to the sum of the antiup and antidown, at $Q^2=20\text{ GeV}^2$, finding it to be $0.52^{+0.18}_{-0.15}$~\cite{Chatrchyan:2013mza}. %1312.6283 
Future high-luminosity studies may help to improve our knowledge of the strangeness. 
In the case of the charm PDFs, there has been a long debate concerning the size of the ``intrinsic'' charm contribution, as first raised in 1980~\cite{Brodsky:1980pb} but still not yet resolved\footnote{We refer interested readers to Ref.~\cite{Brodsky:2015fna} and references within for a review of intrinsic-charm discussions.}.
$c(x)-\overline{c}(x)$ provides an important check of the intrinsic-charm contribution to the proton.
Again, the current experimental data are too inconclusive to discriminate between various proposed QCD models, and future experiments at LHC or EIC could provide useful information in settling this mystery. 

Although there exist a variety of model approaches to treat the structure functions, a nonperturbative approach from first principles, such as lattice QCD (LQCD), provides hope to resolve many of the outstanding theoretical disagreements and provide information in regions that are unknown or difficult to observe in experiments. 
In this work, we will be using the large-momentum effective theory (LaMET) framework~\cite{Ji:2013dva} to provide information on the Bjorken-$x$ dependence of the strange and charm PDFs. 
In the LaMET (or ``quasi-PDF'') approach, time-independent spatially displaced matrix elements that can be connected to PDFs are computed at finite hadron momentum $P_z$. 
A convenient choice for leading-twist PDFs is to take the hadron momentum and quark-antiquark separation to be along the $z$ direction.
On the lattice, we then calculate hadronic matrix elements 
\begin{equation}\label{eq:qPDFME}
h (z,P_z) = \langle P|\bar{\psi}(z)\Gamma W(z,0)\psi(0)|P\rangle,
\end{equation}
where $\psi$ is the quark field (charm and strange in this calculation), 
$|P\rangle$ is the nucleon state in our case, 
$W(z,0)$ is the spacelike Wilson-line product $\left( \prod_n U_z(n\hat{z})\right)$ with $U_z$ a discrete gauge link in the $z$ direction. 
There are multiple choices of operator in this framework that will recover the same lightcone PDFs when the large-momentum limit is taken; in this work, we will use $\Gamma=\gamma^t$ for unpolarized distribution, as suggested in Refs.~\cite{Xiong:2013bka,Radyushkin:2016hsy,Radyushkin:2017cyf,Orginos:2017kos}.
The ``quasi-PDF'' $\tilde{q}(x,P_z)$ are then obtained from a Fourier transformation of the continuum-limit renormalized matrix elements $h^R$
\begin{equation}
\tilde{q}(x,P_z)=\int \frac{dz}{4\pi}e^{ixP_z z} h^R(z,P_z),
\end{equation}
where $x$ is the fraction of momentum carried by the parton relative to the hadron. 
For this first study of these quantities, we will neglect the lattice-spacing and finite-volume dependence.
The quasi-PDF is related to the lightcone PDF at scale $\mu$ in $\overline{\text{MS}}$ scheme through a factorization theorem
\begin{align}
\label{eq:matching}
	\tilde{q}_\psi(x,&P_z,\mu^{\overline{\text{MS}}},\mu^\text{RI},p^\text{RI}_z)=\int_0^1 \frac{dy}{|y|} \times \nonumber\\ &C\left(\frac{x}{y},\left(\frac{\mu^\text{RI}}{p_z^\text{RI}}\right)^2,\frac{yP_z}{\mu^{\overline{\text{MS}}}},\frac{yP_z}{p_z^\text{RI}}\right) 
	q_\psi(y,\mu^{\overline{\text{MS}}}) +...
\end{align} 
where $p_z^\text{RI}$ and $\mu^\text{RI}$ are the momentum of the off-shell strange quark and the renormalization scale in the RI/MOM-scheme nonperturbative renormalization (NPR), 
$C$ is a perturbative matching kernel converting the RI/MOM renormalized quasidistribution to the one in $\overline{\text{MS}}$ scheme used in our previous works~\cite{Chen:2018xof,Lin:2018qky,Chen:2018fwa,Chen:2019lcm}. 
The residual terms, $\mathcal{O}\left(\frac{\Lambda^2_\text{QCD}}{x^2P_z^2},\frac{m_N^2}{P_z^2}\right)$, come from the nucleon-mass correction and higher-twist effects, suppressed by the nucleon momentum. 
Even though there has been multiple PDFs calculation calculated directly at the physical pion mass in recent years~\cite{Lin:2017ani,Chen:2018xof,Lin:2018pvv,Liu:2018hxv,Alexandrou:2018pbm,Alexandrou:2018eet}, only ``connected'' contribution of the PDFs has been studied on the lattice so far.
We refer readers to a recent review article~\cite{Ji:2020ect} that has the most complete summary of the latest $x$-dependent LaMET-related calculations.
This work is the first exploratory study to take on the challenges of the notorious ``disconnected'' contribution, an important next step toward flavor-dependent PDFs from lattice QCD. 

\begin{figure}
\centering	
\includegraphics[width=0.8\linewidth]{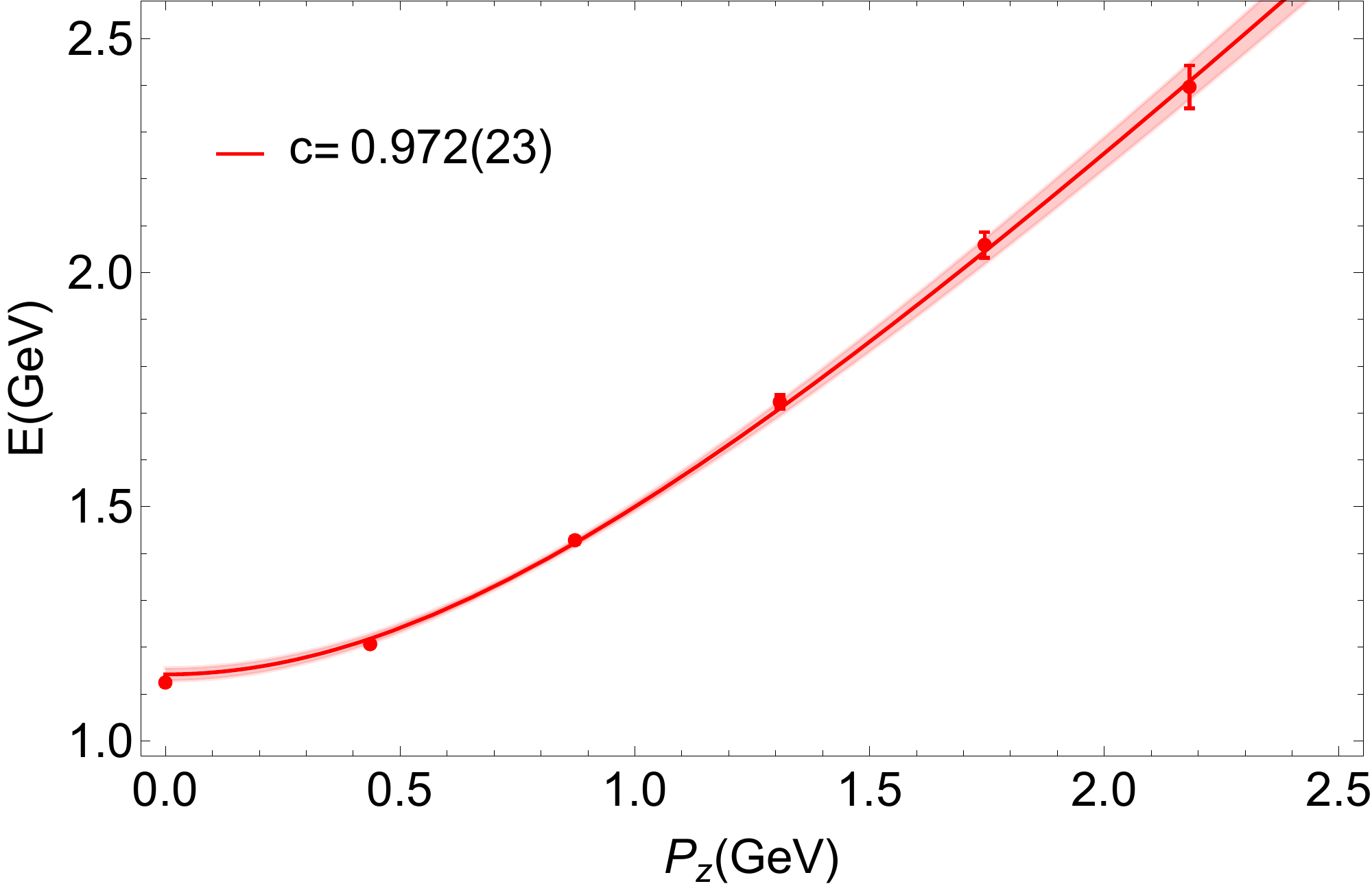}
\caption{The dispersion relation for the nucleon.
The speed of light $c$ from the linear fit $E_0^2=c^2P_z^2+c^4M_N^2$ is slightly smaller than but consistent with 1.}
\label{fig:dispersion_a12}
\end{figure}

%%%%%%%%%%%%%%%%%%%%%%%%%%%%%%%%%%%%%%%%%%%%%%%%%%%%%%%%%%%%%%%%%%%%%%%%%%%%%%%%
We calculate the observables on 898 configurations of the $24^3\times 64$ ensemble with $N_f=2+1+1$ flavors of highly improved staggered quarks (HISQ)~\cite{Follana:2006rc} generated by MILC collaboration ~\cite{Bazavov:2012xda}.
Hypercubic (HYP) smearing~\cite{Hasenfratz:2001hp} is applied to these configurations.
The lattice spacing of this ensemble is $a\approx 0.12$~fm, with $M_\pi \approx 310$~MeV.
The spatial length of this ensemble is approximately 2.88~fm, which gives the $M_\pi^\text{val}L \leq 4.55$.
Past finite-volume studies of nucleon LaMET quasi-PDFs~\cite{Lin:2019ocg} suggest there is negligible likelihood of finite-volume effects in our study here; we will defer study of finite-volume systematics to future works.
The nucleon two-point correlators are constructed with momentum-smeared sources~\cite{Bali:2016lva} to obtain better signal of large-momentum results.
We use momentum-smearing parameters $k=2.9$ and $-2.9$, and calculate $P_z=n_z \frac{2\pi}{L}$ with $|n_z| \in [0,5]$ corresponding to 0 to 2.18~GeV. 
The calculation of two-point correlators includes 57,472 measurements in total. 
At each boost momentum, the nucleon energy is obtained through a two-state fit to the two-point correlator,
$C_\text{2pt}(t)=|\mathcal{A}_0|^2e^{-E_0 t}+|\mathcal{A}_1|^2e^{-E_1 t}+\dots$,
where $E_i$ and $A_i$ are the energy and overlap factor between the lattice nucleon operator and desired state $|i\rangle$, 
and $i=0$ ($i=1$) stands for the ground (excited) state.
Figure~\ref{fig:dispersion_a12} shows the nucleon dispersion relation; we observe that the effective energy of the boosted hadrons grows slightly slower than expected, but the speed of light $c$ is consistent with 1.

\begin{figure}[tb]
\includegraphics[width=0.35\textwidth]{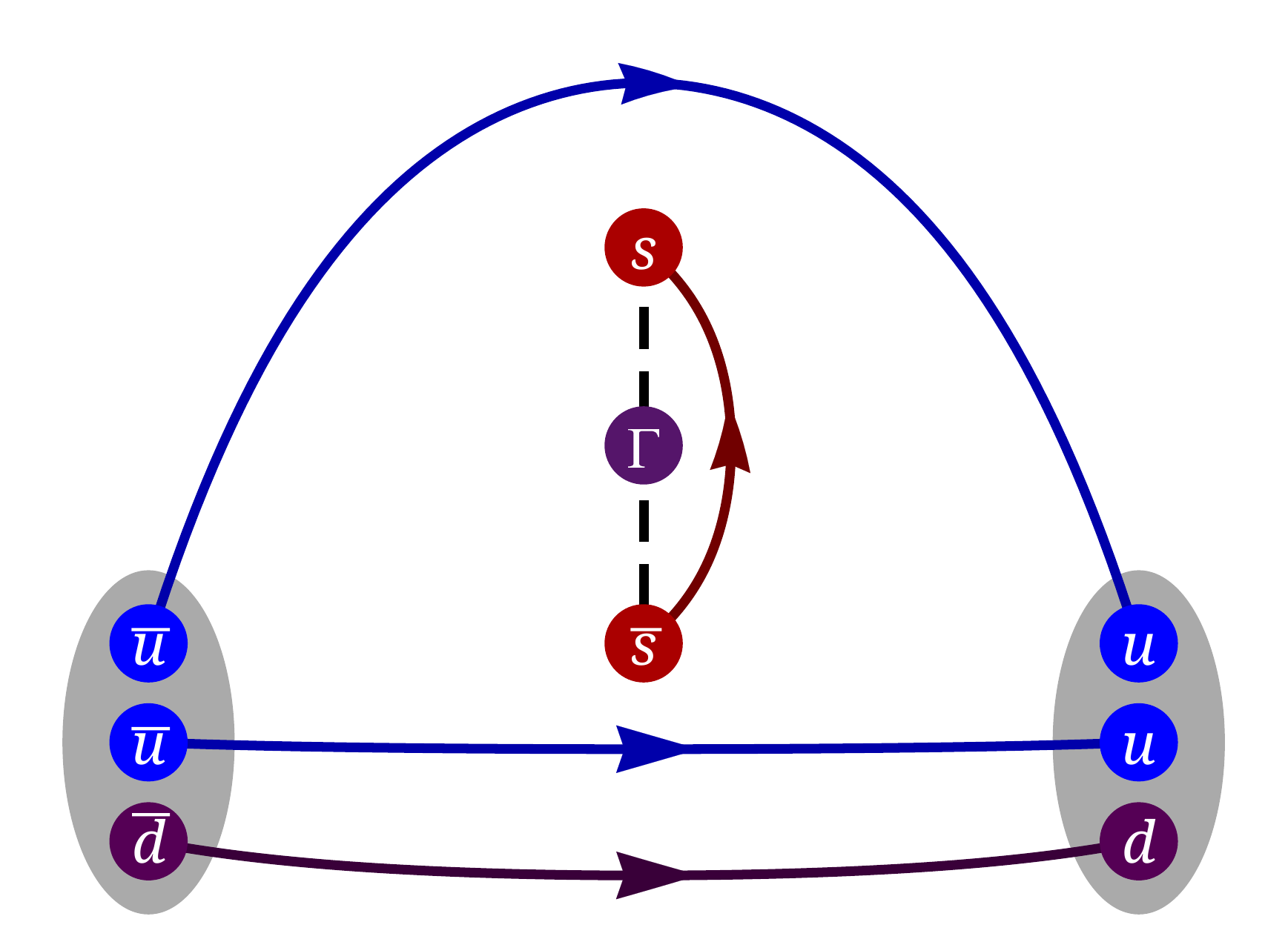}
\caption{
Illustration of the three-point correlation function involving a strange-quark long-link operator, which forms a disconnected diagram.
The dashed line indicates the spatial displacement of the Wilson link with the choice of operator $\Gamma$.
The gray blobs show the nucleon source and sink, separated by $t_\text{sep}$ in Euclidean time direction.
Sea-quarks and gluon interactions, although present in the lattice configurations, are omitted from this schematic diagram. 
\label{fig:con_disc}}
\end{figure}

On the lattice, the only contribution to the strange quasi-PDF matrix elements comes from disconnected quark loops (as shown in Fig.~\ref{fig:con_disc}), calculated as
\begin{equation}
C_{\Gamma}^\text{loop}=\sum_n\Tr\left[(S_{s,c}(n+z\hat{z},n)\prod^{z-1}_{i=0} U_z(n+i\hat{z}))\Gamma\right],
\end{equation}
where $S_{s,c}$ are strange- and charm-quark propagators, $\Gamma=\gamma_t$ gives the unpolarized quasi-PDF, and $n$ indexes over lattice sites.
One of the main challenges to finding the nucleon strange and charm content is calculating the computationally expensive and statistically noisy disconnected diagrams.
We calculate the disconnected diagrams using a stochastic estimator with noise sources accelerated by a combination of the truncated-solver method~\cite{Collins:2007mh, Bali:2009hu}, the hopping-parameter expansion~\cite{Thron:1997iy, Michael:1999rs} and the all-mode--averaging technique~\cite{Blum:2012uh}. 
These methods of calculating quark-line disconnected contributions have proven to be useful in extracting the up, down and strange contributions to the nucleon tensor charges and setting an upper bound for BSM that is dominated by quark EDM~\cite{Bhattacharya:2015wna,Bhattacharya:2015esa}. 
For the disconnected loop in this calculation, we have a total of 3,592,000  low-precision (LP) measurements ($N_\text{LP}=4000$ for each configuration) and 71,840 high precision (HP) measurements. 
Once we obtain the strange/charm loop, we can construct the strange/charm nucleon three-point correlators ($C_{\text{3pt}}$) by combining it with the two-point correlator ($C_{\text{2pt}}$):
\begin{align}
\label{eq:combine_3pt}
 C_{\text{3pt}}(t,t_\text{sep})=&\langle\left( C_{\text{2pt}}(t_\text{src},t_\text{sep})-\langle C_{\text{2pt}}(t_\text{src},t_\text{sep})\rangle\right)
 \nonumber\\
 \cdot&\left(C_{\gamma_t}^\text{loop}(t+t_\text{src})-\langle C_{\gamma_t}^\text{loop}(t+t_\text{src})\rangle\right)\rangle_{t_\text{src}},
\end{align}
where $t_\text{src}$ and $t_\text{sep}$ are the source location and source-sink separation, respectively.

\begin{figure}[tb]
\includegraphics[width=0.8\linewidth]{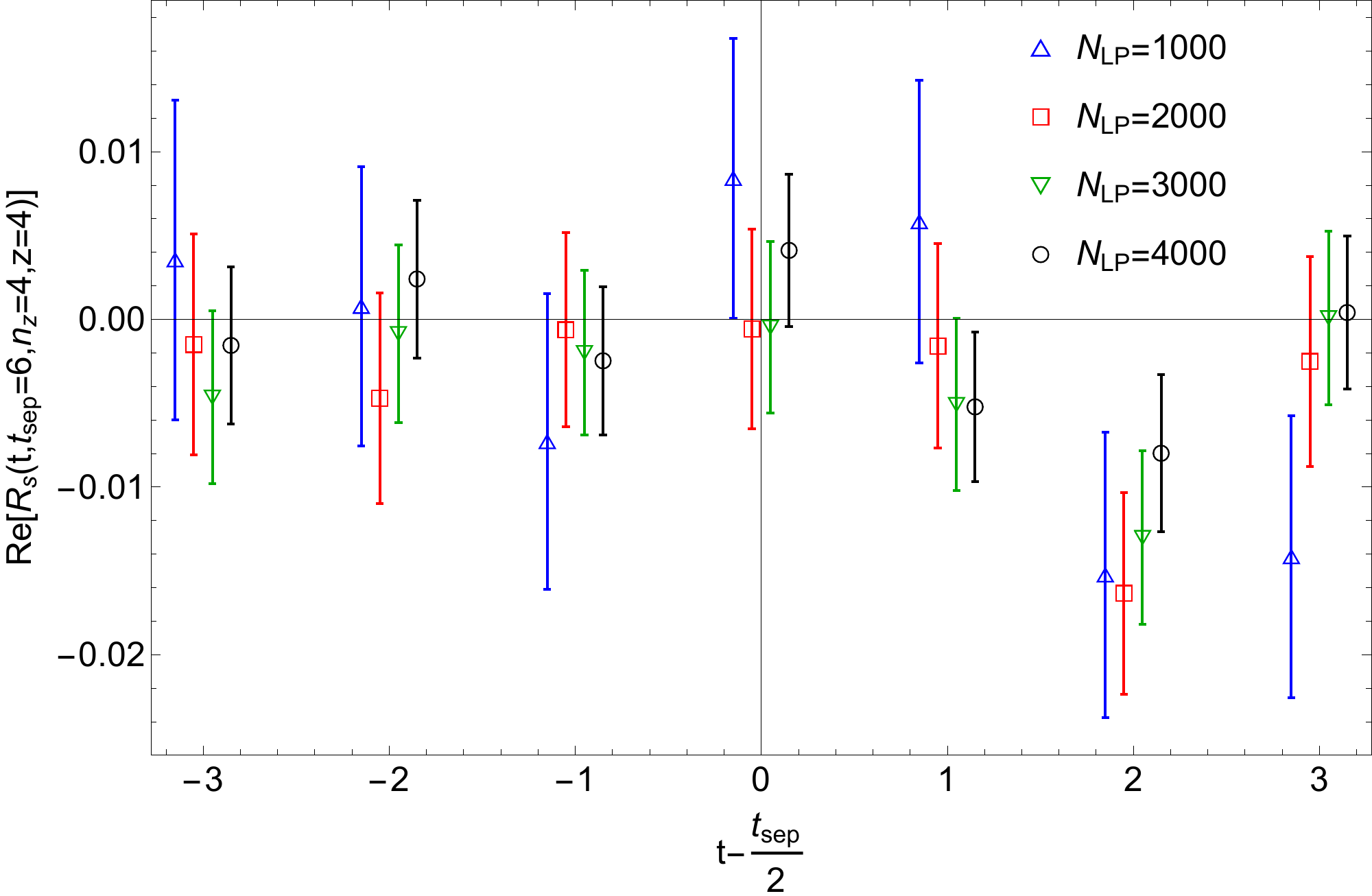}
\includegraphics[width=0.8\linewidth]{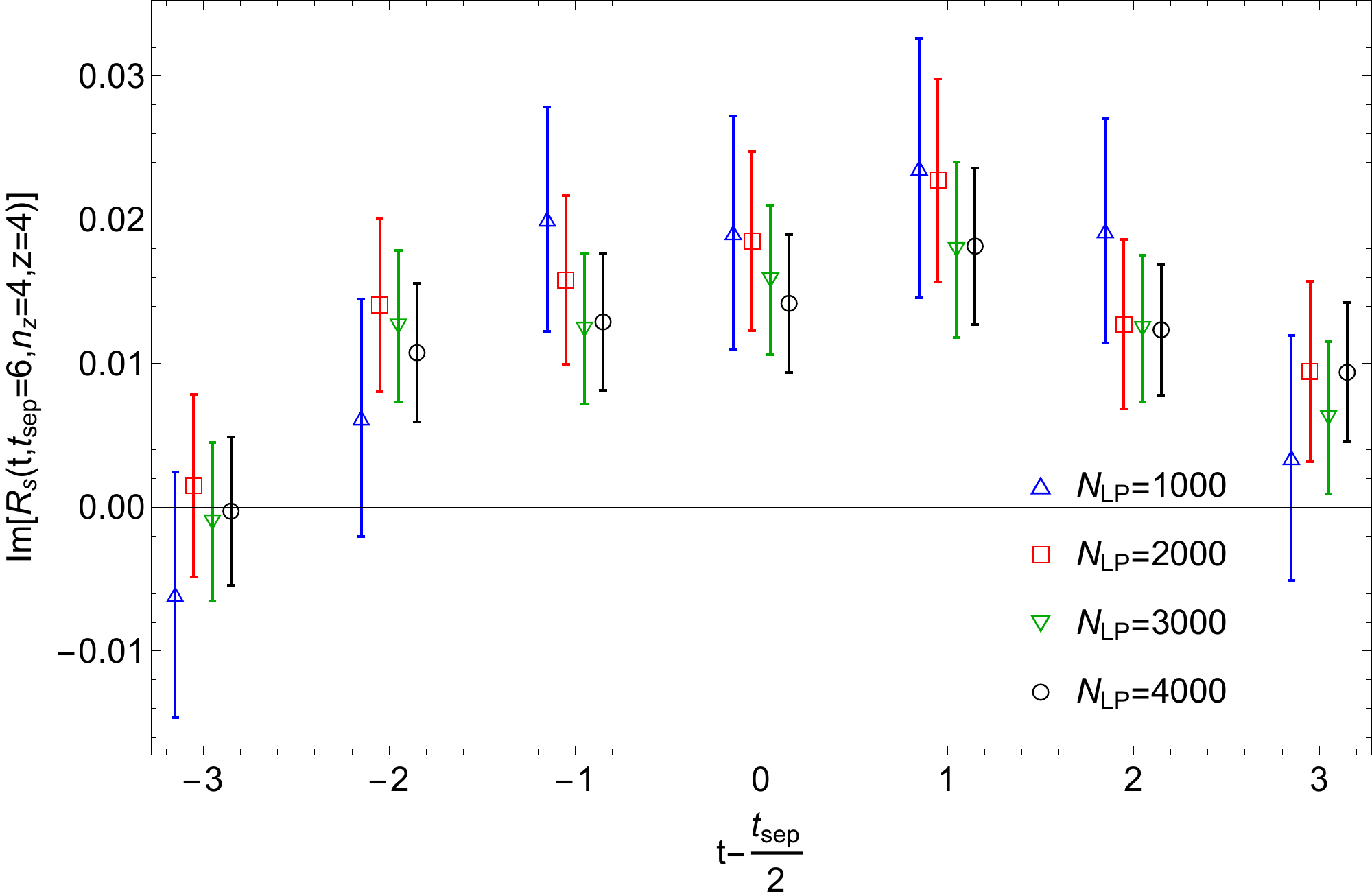}
\caption{
Example ratio plot for real (top) and imaginary (bottom) strange nucleon matrix elements from $n_z=4$, $z=2$, $t_\text{sep}=6$ as function of insertion time $t$, centered by half of $t_\text{sep}$. 
Different data points indicate the value of $N_\text{LP}\in\{1000,2000,3000,4000\}$ noise sources used in the calculation (with slightly shift in $t$ to make the data points visible).
The error is effectively reduced by using larger $N_\text{LP}$.
\label{fig:saturation}}
\end{figure}

To obtain the ground-state nucleon strange matrix elements, we fit the three-point correlators, which are expanded in energy eigenstates as
\begin{align}
\label{eq:fit_form}
C_\text{3pt}(t_\text{sep},t)=&|\mathcal{A}_0|^2\langle0|O|0\rangle e^{-E_0 t_\text{sep}}\nonumber\\
&+\mathcal{A}_0\mathcal{A}_1^*\langle0|O|1\rangle e^{-E_0 t}e^{-M_1 (t_\text{sep}-t)}\nonumber\\
&+\mathcal{A}^*_0\mathcal{A}_1\langle1|O|0\rangle e^{-E_0 (t_\text{sep}-t)}e^{-E_1t}\nonumber\\
&+|\mathcal{A}_1|^2\langle1|O|1\rangle e^{-E_1 t_\text{sep}}+\dots
\end{align}
where 
$\langle i'|O|i\rangle$ indicates matrix elements for the ground-state ($i=i'=0$) or excited states ($i'=1$).
The ground-state matrix element we want to obtain for the LaMET operator is $\langle 0|O|0\rangle$, which could be approximated by the ratio
\begin{equation}
R_V(t_\text{sep},t)=\langle C_\text{3pt}(t_\text{sep},t)\rangle/\langle C_\text{2pt}(t_\text{sep})\rangle
\end{equation}
if the excited-state contamination in the data were small.  
Figure~\ref{fig:saturation} shows one example of real and imaginary ratio plots for the strange nucleon correlators at $n_z=4$, $z=2$ at $t_\text{sep}=6$ using different numbers of the low-precision sources, varying from 1000 to 4000. 
We find the statistical errors consistently decrease as the $N_\text{LP}$ increases, approximately scaling as $\frac{1}{\sqrt{N_\text{LP}}}$.  

\begin{figure}[tb]
\includegraphics[width=0.8\linewidth]{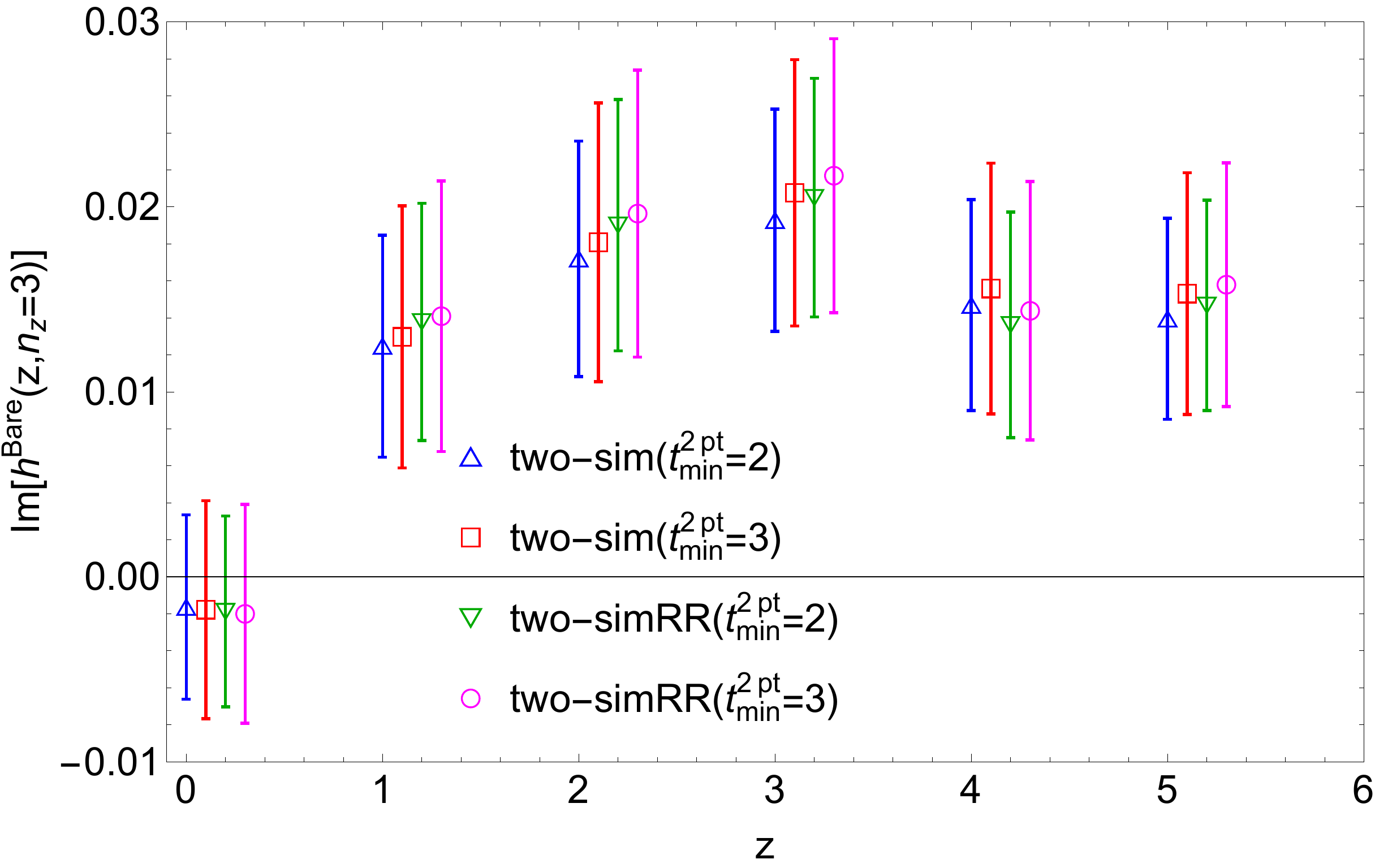}
\caption{
Comparison of the imaginary part of the strange quasi-PDF bare matrix elements obtained from different fit strategies at $n_z=3$.
\label{fig:fit_stability}}
\end{figure}

\begin{figure}[tb]
\includegraphics[width=0.8\linewidth]{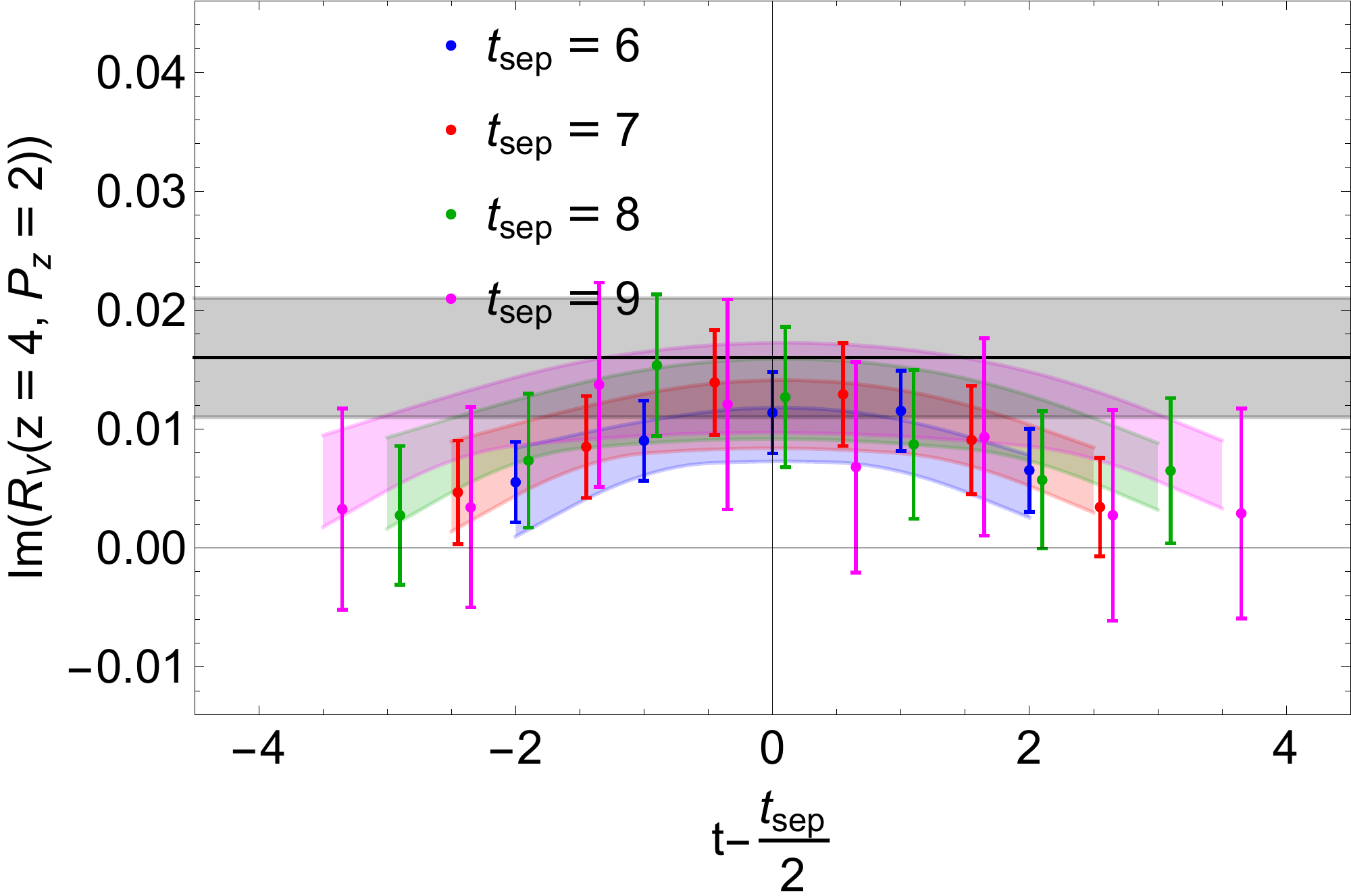}
\includegraphics[width=0.8\linewidth]{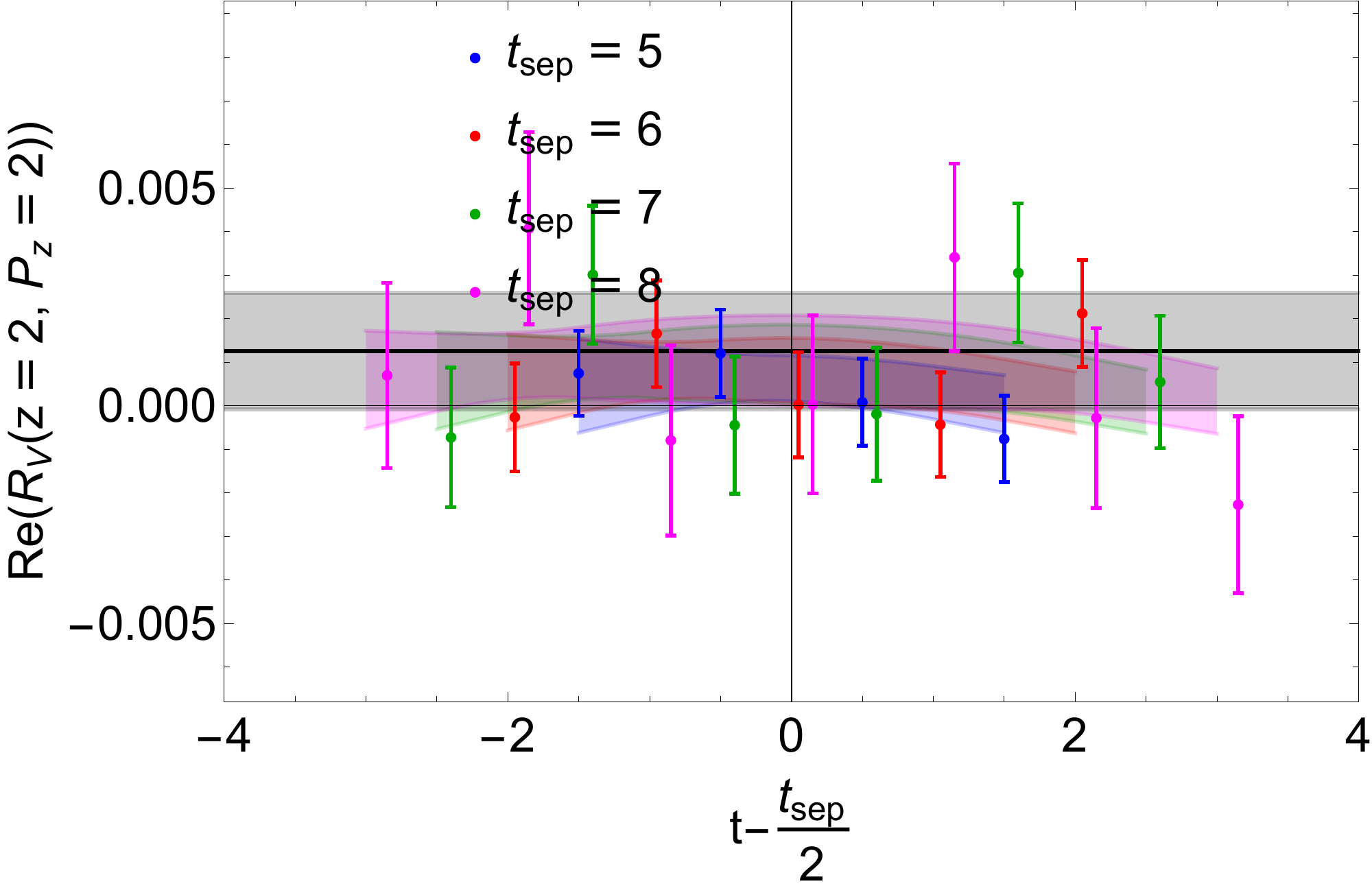}
\caption{
Ratio plots for the imaginary part of strange quasi-PDF at $n_z=2$, $z=4$ (top) and the real part of charm quasi-PDF at $n_z=2$, $z=2$ (bottom).
The ratios are plotted as data points with error bars and the fitted results are plotted in colored bands.
The gray band is the ground-state matrix elements obtained from the fit.
\label{fig:ratio}}
\end{figure}

We check the stability of the fit results using different strategies: fitting the two-point correlators of $t\in[2,10]$ to the first two states in Eq.~\eqref{eq:fit_form} and the three-point correlators within $t_\text{sep}\in[6,9]$ and $t\in[1,t_\text{sep}-1]$ to the first three terms (two-sim) or four terms (two-simRR).
An example of the fitted bare matrix elements is shown in Fig.~\ref{fig:fit_stability}.
The fit results are consistent among different strategies.
In the remaining part of the paper we adopt the two-sim strategy and $t^\text{2pt}_\text{min}=3$ for the fits.
Two selected ratio plots with fit results for the imaginary part of strange quasi-PDF at $n_z=2$, $z=4$ and the real part of charm quasi-PDF at $n_z=2$, $z=2$ are shown in Fig.~\ref{fig:ratio}.
The ratios calculated from correlators are presented as data points with error bars, while the fitted results are plotted as colored bands.
The gray band is the ground-state matrix element we extract from the fit results.
We observe that the real ratios are consistent with zero, and the imaginary ratios are consistent with the fitted results.

%%%%%%%%%%%%%%%%%%%%%%%%%%%%%%%%%%%%%%%%%%%%%%%%%%%%%%%%%%%%%%%%%%%%%%%%%%%%%%%%
We then apply the renormalization factors to the bare matrix elements, in order to make comparisons with other results. 
We adopt nonperturbative renormalization (NPR) in RI/MOM scheme, the same strategy as in past works~\cite{Stewart:2017tvs,Chen:2017mzz}, by imposing 
\begin{multline}\label{hRx}
Z(p^\text{RI}_z, \mu^\text{RI}) = \\
\left.\frac{\Tr[\slashed p \sum_s \langle p,s| \bar\psi_f(\lambda \tilde n) \slashed{\tilde n_t} W(\lambda\tilde n,0) \psi_f(0)|p,s\rangle]}
{\Tr[\slashed p  \sum_s \langle p,s| \bar\psi_f(\lambda \tilde n) \slashed{\tilde n_t} W(\lambda\tilde n,0) \psi_f(0) |p,s\rangle_\text{tree}]} \right|_{\tiny\begin{matrix}p^2=-\mu_\text{RI}^2 \\ \!\!\!\!p_z=p^\text{RI}_z\end{matrix}}.
\end{multline}
We use the NPR factors in RI/MOM scheme calculated from Ref.~\cite{Lin:2020ssv}  to obtain the renormalized matrix elements:
$h^R(z,p_z^\text{RI}, \mu^\text{RI}) = Z^{-1}(p_z^\text{RI}, \mu^\text{RI}) h(z)$; throughout this work, we will fix the scales to $p_z^\text{RI}=0$, $\mu^\text{RI}=2.3$~GeV.

\begin{figure}
\centering
\includegraphics[width=0.8\linewidth]{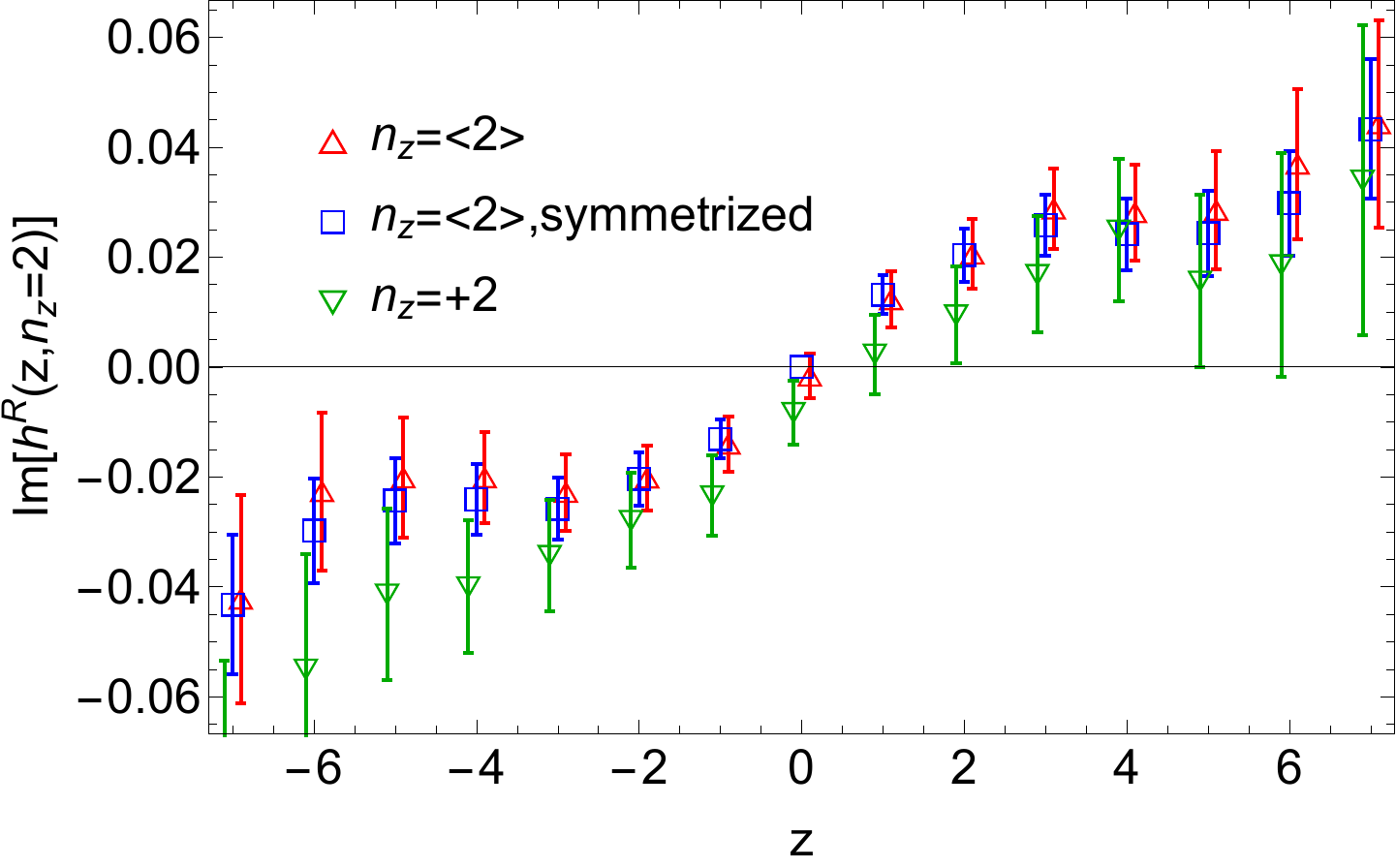}
\caption{\label{fig:average_dir_a12}
Comparison of the imaginary parts of the renormalized nucleon strange-PDF matrix elements as functions of $z$ using different momentum-smearing parameters and symmetrization methods.
The results are obtained at $M_\pi \approx 310$~MeV and nucleon boost momentum $P_z=0.87$~GeV.}
\end{figure}

To confirm that we are observing signal, given the small magnitude of the matrix elements, we also check whether averaging the results of the nucleon momentum in opposite directions of the smearing momentum parameter improves the signal (which also preserve rotational symmetry in the data). 
Figure~\ref{fig:average_dir_a12} shows example renormalized fitted imaginary matrix elements at one of the boost momenta, $P_z=0.88$~GeV as a function of the dimensionless parameter $zP_z$.
It shows that the data from a single positive and the average of two opposite momentum-smearing results are consistent within statistical errors.
We also find that averaging over opposite directions effectively increases the statistics by a factor around 2.
Furthermore, to satisfy the requirement that the quasi-PDF is real in momentum space, the matrix elements in coordinate space must satisfy $h(z)=h(-z)^*$; this is also observed in our data. 
We can then utilize this relationship to further improve the signal in our matrix elements.
This symmetrization also improves the statistics of our matrix elements, as shown in Fig.~\ref{fig:average_dir_a12}.
For the rest of this paper, we only present the matrix elements that have been averaged over momentum-smearing and symmetrized across negative link lengths.

\begin{figure}[t!]
\centering
\includegraphics[width=0.8\linewidth]{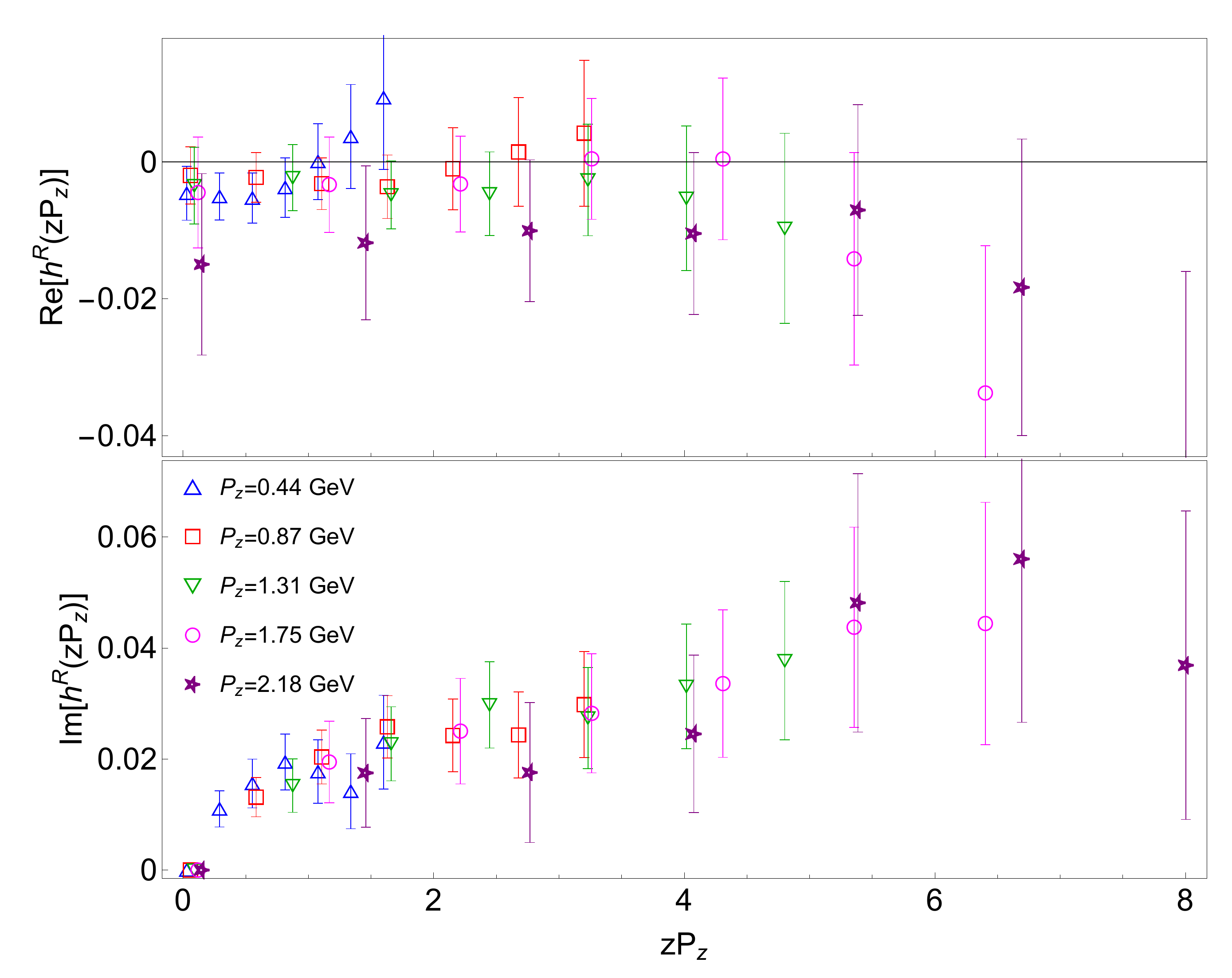}
\includegraphics[width=0.8\linewidth]{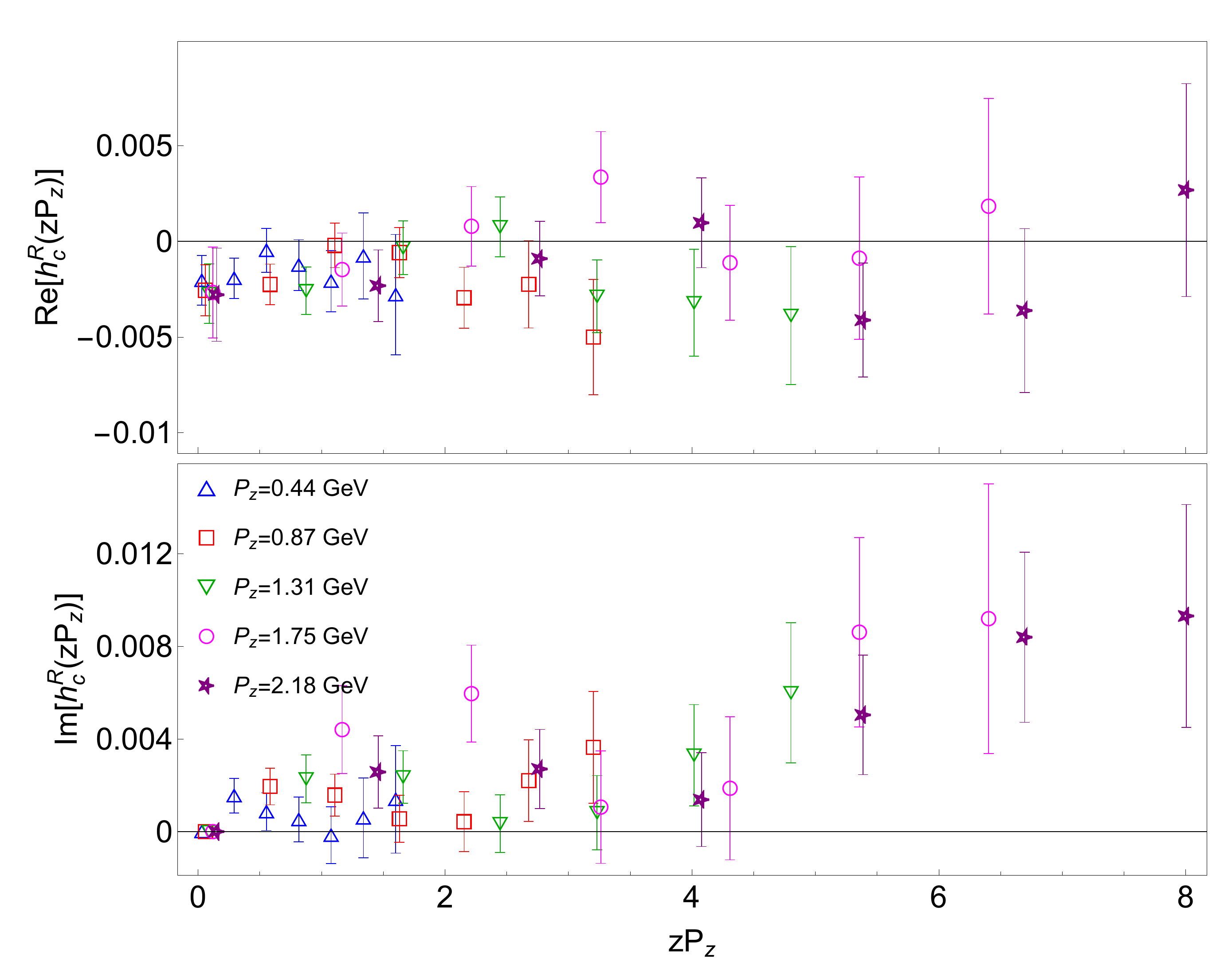}
\caption{\label{fig:lamet_zpz_a12_msu_light}
RI/MOM renormalized strange (top) and charm (bottom) nucleon quasi-PDF matrix element %at $\mu^\text{RI}=2.3 \text{GeV}$ 
as a function of $zP_z$ at $P_z\in [0.44,2.18]$ at $M_\pi\approx 310$~MeV. 
The real matrix elements are consistent with zero within 2 sigma; suggesting quark-antiquark symmetry.}
\end{figure}

Figure~\ref{fig:lamet_zpz_a12_msu_light} shows the symmetrized and renormalized strange and charm quasi-PDF matrix elements for the nucleon of $M_\pi \approx 310$~MeV.
We find that the matrix elements calculated at different boost momenta can have small discrepancies, but they are consistent with each other at large momentum and seem to be approaching a universal curve.
The real quasi-PDF matrix elements are consistent with zero at 95\% confidence level for most $zP_z$ points, indicating that the quark-antiquark asymmetries for both strange and charm are likely very small.
The imaginary matrix elements from strange quasi-PDFs are about one order of magnitude larger than those of charm, which is consistent with the magnitudes obtained from the global fitting of strange and charm PDFs.  
The same quantities are also calculated for the nucleon at the SU(3) point, where the light-quark masses are equal to the physical strange-quark mass.
The matrix elements have similar behavior but have better signals than our light nucleon results.

\begin{figure}[tb]
\centering
\includegraphics[width=0.8\linewidth]{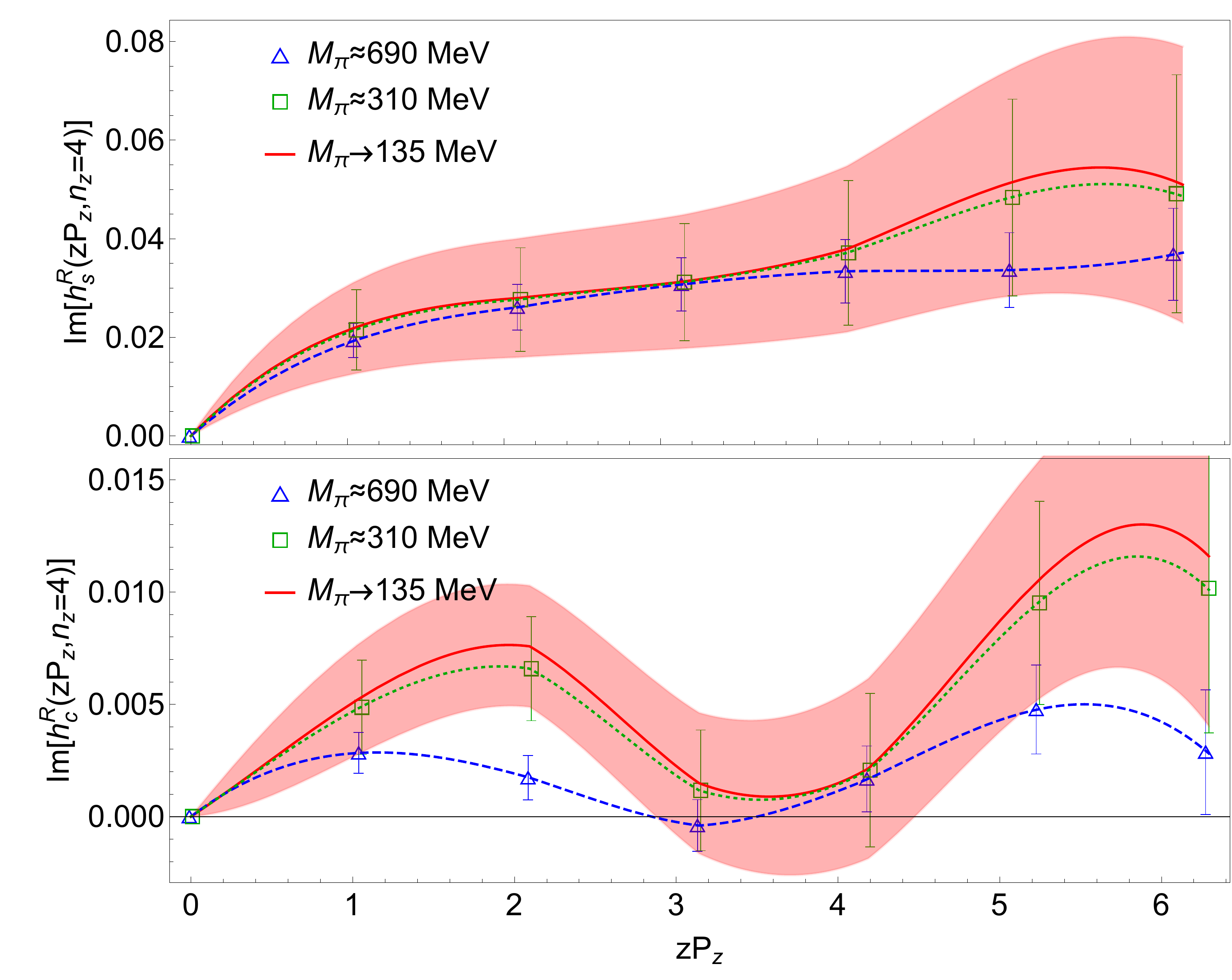}
\caption{Extrapolation to physical pion mass for the imaginary part of strange (top) and charm (bottom) MEs at $n_z=4$.
The extrapolated results are very close to the $M_\pi \approx 310$~MeV results.
The charm distribution is much smaller, thus noisier with larger relative errors.
The band indicates the extrapolated matrix elements at the physical pion mass. 
}
\label{fig:extrapolation}
\end{figure}

With the two mass points $M_\pi \approx 310$~MeV and $M_\pi \approx 690$~MeV, we perform a naive chiral extrapolation with the form
$h^R(M_\pi) = h^R_\text{phys} + c_1 (M_\pi^2-M_{\pi,\text{phys}}^2)$ 
to estimate the matrix elements at physical pion mass $M_{\pi,\text{phys}} \approx 135$~MeV. 
We show example extrapolated results for the imaginary strange matrix element at boost momentum $P_z=1.76$~GeV in the top panel of Fig.~\ref{fig:extrapolation}, along with the results for the charm matrix elements (bottom panel).
Both extrapolated matrix elements are very close to those from the 310-MeV calculation. 
In the strange case, we observe a small pion-mass dependence between the 310 and 690-MeV results at large $P_z$.
This is understandable, since as $P_z$ increases, the system scale is dominated by energy which mainly contribute by $P_z$ when $P_z > M_N$. 
The charm matrix elements are smaller and show signs of oscillating statistically as a function of $zP_z$, 
whereas the strange distribution shows a tendency to keep growing as $zP_z$ increases.

\begin{figure}[tb]
\centering	
\includegraphics[width=0.9\linewidth]{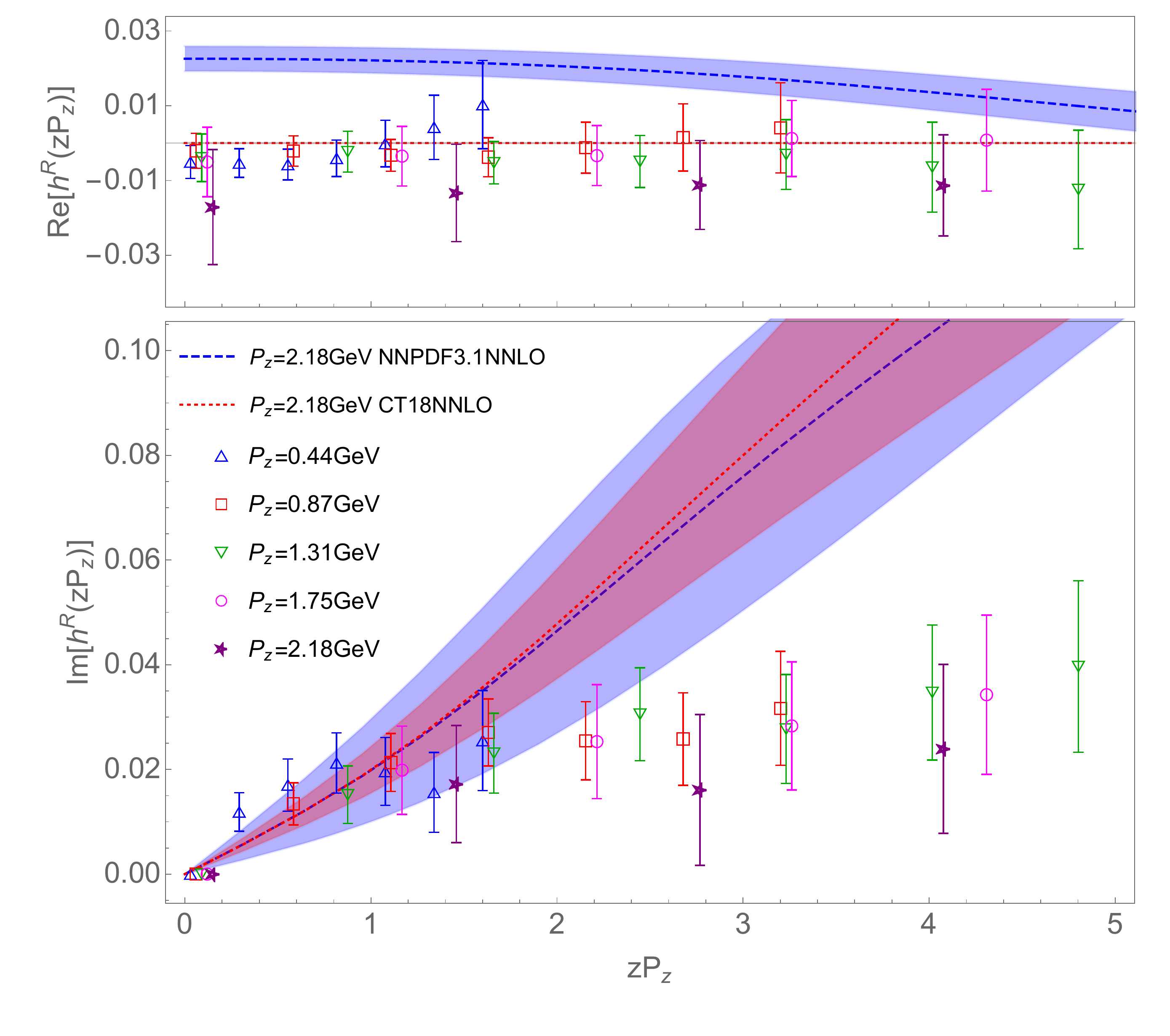}
\caption{\label{fig:lamet_global_fit_msu}
The real (top) and  imaginary (bottom) parts of the strange quasi-PDF matrix elements in coordinate space from our calculations at physical pion mass with $P_z \in [0.44, 2.18]$~GeV, along with those from CT18 and NNPDF NNLO in RI/MOM renormalized scale of 2.3~GeV.
The CT18 analysis assumes $s(x)=\bar{s}(x)$, so their results are exactly zero after matching and Fourier transformation.  
Our real matrix elements are all consistent with zero, supporting strange-antistrange symmetry, while our imaginary ones are smaller at large $zP_z$.}
\end{figure}

We compare our extrapolated matrix-element results with those obtained from global fitting of strange PDFs from CT18NNLO~\cite{Hou:2019efy} and the NNPDF3.1NNLO~\cite{Ball:2017nwa} at 2~GeV in $\overline{\text{MS}}$ scheme provided by LHAPDF~\cite{Buckley:2014ana}; we match these to RI/MOM renormalization at $\mu^\text{RI}= 2.3$~GeV with $P_z=2.18$~GeV quasi-PDF matrix elements, using Eq.~\ref{eq:matching}.
The real matrix elements are proportional to the integral of the difference between strange and antistrange ($\int dx \left(s(x)-\bar{s}(x)\right) \cos(xzP_z)$); our results, as shown in Fig.~\ref{fig:lamet_global_fit_msu}, are consistent with zero at most $zP_z$, suggesting a symmetric $s-\bar{s}$ distribution. 
The CT18NNLO PDFs assumes a symmetric $s-\bar{s}$ distribution, so are exactly zero under the transformation with the renormalization scale we used in this work, consistent with our findings in this work.
The imaginary matrix elements are proportional to $\int dx \left(s(x)+\bar{s}(x)\right) \sin(xzP_z)$. 
The pseudo-PDF matrix elements from both CT18 and NNPDF are consistent with our results within 2 standard deviations up to $zP_z \approx 3$, and deviate from our results at large $zP_z$, suggesting deviations at moderate to small-$x$ in the PDFs.
However, the matching kernel we used in this work is only valid for nonsinglet structure, such as $s(x)-\bar{s}(x)$; it is not complete for $s(x)+\bar{s}(x)$.
To properly account for the full strange PDFs, we will need the full light-flavor contribution, as well as the gluon one, to apply the full matching kernel with mixing.
Future study will be necessary to discern the full strange PDF structure from lattice calculations.  

Similarly, we compare the charm results with the global-fit PDFs in Fig.~\ref{fig:lamet_global_fit_msu}.
Note that CT18 and NNPDF3.1 both assume $c(x)=\bar{c}(x)$; therefore, both of them have vanishing real matrix elements, which is consistent with the our real matrix elements for the charm quasi-PDF. 
Our imaginary charm matrix elements have much smaller magnitude than the strange, a similar strange-charm relation also observed by global PDF fitting, such as CT18 and NNPDF. 
The charm PDF errors from global fits are significantly different because the CT18 charm PDF is generated by perturbatively evolving from light-quark and gluon distributions at $Q_0 = 1.3$~GeV while NNPDF numerically fitted the charm distribution.
Our imaginary matrix elements are close to zero at small $zP_z$; at large $zP_z$ they are about a factor of 5 smaller than the strange ones and are within the bounds of the NNPDF results.

\begin{figure}[tb]
\centering	
\includegraphics[width=0.9\linewidth]{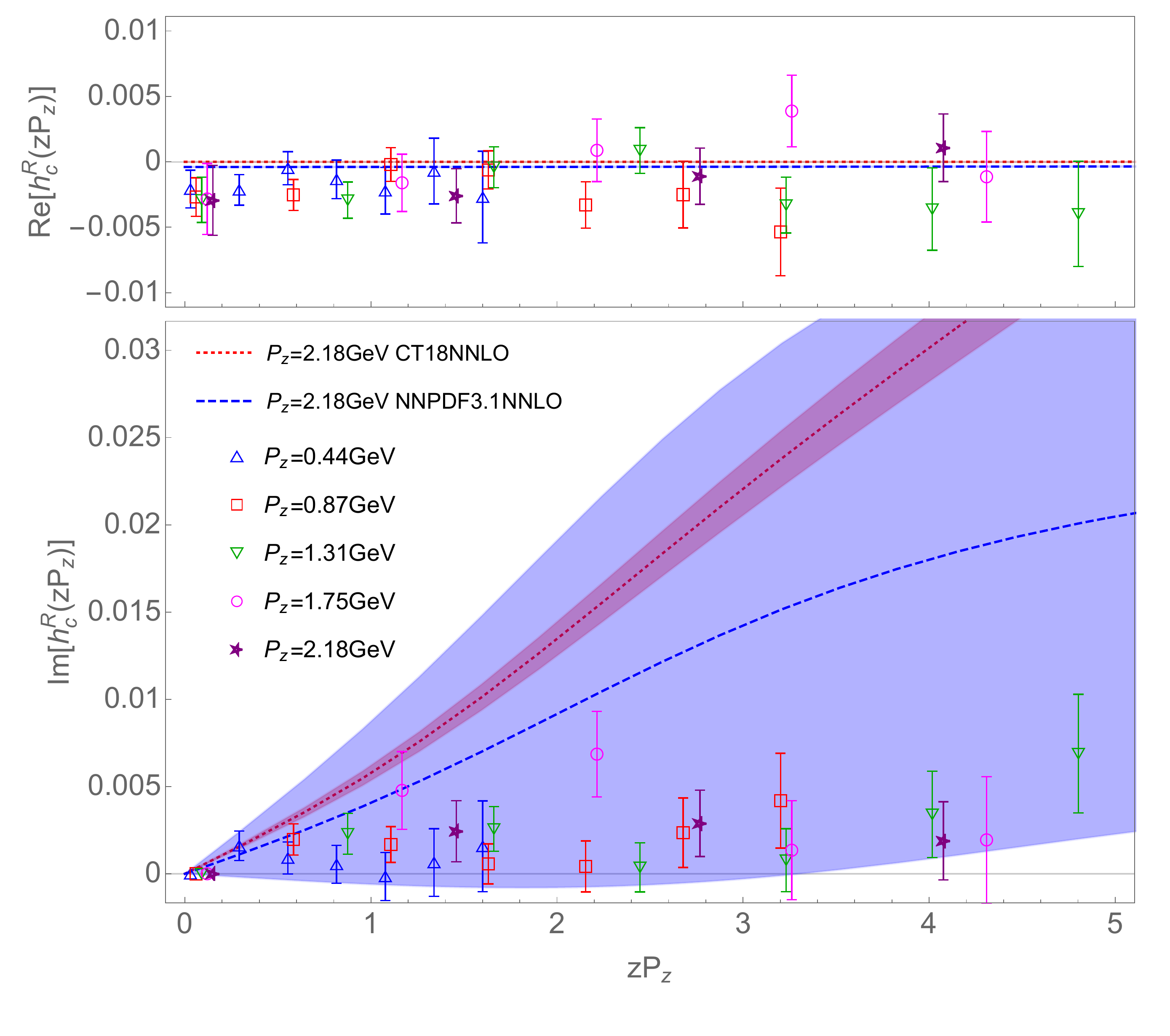}
\caption{\label{fig:lamet_global_fit_msu_c}
The real (top) and  imaginary (bottom) parts of the charm  
quasi-PDF matrix elements in coordinate space derived from global-fit PDFs compared with our renormalized nucleon quasi-PDF MEs at $P_z \in [0.44, 2.18]$~GeV.  
}
\end{figure}

In this work, we made the first lattice-QCD calculations of the strange and charm parton distributions using LaMET (also called ``quasi-PDF'') approach on a single 2+1+1-flavor HISQ ensemble with physical strange and charm masses and heavier-than-physical light-quark mass (resulting in a 310-MeV pion).
We found that our renormalized real matrix elements are zero within our statistical errors for both strange and charm, supporting the strange-antistrange and charm-anticharm symmetry assumptions commonly adopted by most global PDF analyses.
Our imaginary matrix elements are proportional to the sum of the quark and antiquark distribution, and we clearly see that the strange contribution is about a factor of 5 or larger than charm ones.
They are consistently smaller than those from CT18 and NNPDF, possibly due to missing the contributions from other flavor distributions in the matching kernel.
Higher statistics will be needed to better constrain the quark-antiquark asymmetry.
A full analysis of lattice-QCD systematics, such as finite-volume effects and discretization, is not yet 
included, and plans to extend the current calculations are underway.

%%%%%%%%%%%%%%%%%%%%%%%%%%%%%%%%%%%%%%%%%%%%%%%%%%%%%%%%%%%%%%%%%%%%%%%%%%%%%%%%
{\bf Acknowledgments: }
We thank the MILC Collaboration Collaboration for sharing the lattices used to perform this study. The LQCD calculations were performed using the Chroma software suite~\cite{Edwards:2004sx} with the multigrid solver algorithm~\cite{Babich:2010qb,Osborn:2010mb}. 
This research used resources of 
%1.
the National Energy Research Scientific Computing Center, a DOE Office of Science User Facility supported by the Office of Science of the U.S. Department of Energy under Contract No. DE-AC02-05CH11231 through ERCAP; 
%2. 
facilities of the USQCD Collaboration, which are funded by the Office of Science of the U.S. Department of Energy, 
Extreme Science and Engineering Discovery Environment (XSEDE), which is supported by National Science Foundation grant number ACI-1548562, 
%3. 
and supported in part by Michigan State University through computational resources provided by the Institute for Cyber-Enabled Research (iCER). 
HL and RZ are supported by the US National Science Foundation under grant PHY 1653405 ``CAREER: Constraining Parton Distribution Functions for New-Physics Searches''.
BY is supported by the U.S. Department of Energy, Office of Science, Office of High Energy Physics under Contract No. 89233218CNA000001 and by the Los Alamos National Laboratory (LANL) LDRD program. 
BY also acknowledges support from the U.S. Department of Energy, Office of Science, Office of Advanced Scientific Computing Research and Office of Nuclear Physics, Scientific Discovery through Advanced Computing (SciDAC) program.
\\
\\
\\
\\

\end{document}